\documentclass[aps,twocolumn,pra]{revtex4}
\usepackage{psfrag,graphicx}
\def\be{\begin{equation}}
\def\ee{\end{equation}}
\def\bea{\begin{eqnarray}}
\def\eea{\end{eqnarray}}
\def\no{\nonumber}

\def\e{\epsilon}

\begin{document}
\title{Unbreakable $\mathcal{PT}$ symmetry of exact solitons supported
by transversally modulated nonlinearity acting as a
psuedopotential}
\author{Thokala Soloman Raju}
\email{soloman@iisertirupati.ac.in}
\address{Indian Institute of Science Education and Research
(IISER) Tirupati, Andhra Pradesh 517507, India
  }
\author{Tejaswi Ashok Hegde}
\address{
Department of Physics, Karunya University,~ Coimbatore 641 114,
India}
  \author{C.N. Kumar}
  \address{Department of Physics, Panjab University, Chandigarh 160 014, India}
\begin{abstract}
We demonstrate analytically and numerically the existence of exact
solitons in the form of double-kink and fractional-transform,
supported by a symmetric transversally modulated defocusing
nonlinearity acting as a psuedopotential combined with an
antisymmetric gain-loss profile. We explicate here that the
$\mathcal{PT}$ symmetry is never broken in the dynamical system
under study, even in the absence of any symmetric modulation of
linear refractive-index, and the transversally modulated
defocusing nonlinearity comes in the way as a requirement to
establish the ensuing $\mathcal{PT}$ symmetry.
\end{abstract}
\maketitle

\section{Introduction}
Much attention has been paid to the study of light propagation in
parity-time ($\mathcal{PT}$) symmetric optical media. As demanded
by quantum mechanical notion that every physical observable
associated with a real spectrum must be Hermitian, these
$\mathcal{PT}$-symmetric optical structures deliberately exploit
this quantum mechanical notion of parity and time reversal
symmetry of non-Hermitian Hamiltonian. Indeed, a non-Hermitian
Hamiltonian having unbroken $\mathcal{PT}$-symmetry possesses
entirely real and positive energy eigenvalues and may describe a
physically viable system without violating any of the postulates
of quantum mechanics.

In order to implement these $\mathcal{PT}$-symmetric settings in
optics, it amounts to combine spatially symmetric refractive-index
landscapes with mutually balanced spatially separated gain and
loss. These ideas were proposed in
Refs.~\cite{muga,musslimani,berry,klaiman,longhi1,longhi2,kottos,miri}
and demonstrated in Refs.~\cite{guo,segev,regen}. Subsequently,
these works had drawn a great deal of attention to models of
optical systems featuring the $\mathcal{PT}$-symmetry \cite{15}. A
majority of such models actually include Kerr nonlinearity, and
they are modeled by nonlinear Schr\"odinger equation with a
complex potential, whose real and imaginary parts are said to be
spatially even and odd. Some of the interesting situations occur
when the underlying evolution equations contain only nonlinear
$\mathcal{PT}$-symmetric terms \cite{kartashov1,kartashov2} or
mixed linear-nonlinear lattices \cite{kivshar,he,he2}.

In this paper, we propose to explicate the existence of exact
chirped solitons in the form of double-kink and
fractional-tarnsform supported by a symmetric transversally
modulated defocusing nonlinearity acting as a psuedopotential
combined with an antisymmetric gain-loss profile. We demonstrate
here that the $\mathcal{PT}$ symmetry is never broken in the
dynamical system under study, even in the absence of any symmetric
modulation of linear refractive-index. For a homogeneous
nonlinearity, the $\mathcal{PT}$ symmetry is always broken, but
for the transversally modulated defocusing nonlinearity we
demonstrate analytically that the ensuing $\mathcal{PT}$ symmetry
is never broken. We corroborate this fact for two specific
examples of nonlinearity and the gain-loss profiles. Recently, for
a special case of $\mathcal{PT}$-symmetric Hamiltonian system, the
phenomenon of unbreakable symmetry was demonstrated for a dimer
\cite{barashenkov}. In the present work, we show that in the
presence of the nonlinearity modulation, the symmetry is said to
become unbreakable, as it holds at arbitrarily large strengths of
the balanced gain and loss. Furthermore, we observe that the chirp
associated with each of these exact solutions is dependent on the
intensity of the wave quadratically. We begin our analysis by
considering optical wave propagation in self-defocusing Kerr
nonlinear $\mathcal{PT}$ symmetric potential. In this case, the
beam evolution is governed by the following normalized nonlinear
Schr\"odinger equation, \be\label{3a1}
i\frac{\partial\psi}{\partial
z}=-\frac{1}{2}\frac{\partial^{2}\psi}{\partial\eta^{2}}+\sigma(\eta)\psi|\psi|^{2}+iG(\eta)\psi.
\ee where $\psi$ is the scaled amplitude and is proportional to
the electric field envelope; $z$ and $\eta$ are the propagation
distance normalized to the diffraction length $kx_0^2$ and
transverse coordinate normalized to the characteristic transverse
scale $x_0$ respectively; the function $\sigma(\eta)>0$, which is
assumed to be even, describes the profile of a self-defocusing
nonlinearity; and the function $G(\eta)$, assumed to be odd,
stands for the gain-loss profile. Recently, Kartashov et al
\cite{kartashov3} adopted a different variety of nonlinearity and
gain-loss profiles and examined unbreakable $\mathcal{PT}$
symmetry of solitons numerically and analytically. Motivated by
this work and others \cite{kartashov4,kartashov5}, we explicate
here, this phenomenon of unbreakable $\mathcal{PT}$ symmetry for
two special cases of nonlinearity function and the gain-loss
profiles: Case(i):--- $\sigma(\eta)=\frac{m^{2}{\rm
sinh}^{2}(n\eta)}{\epsilon +{\rm sinh}^{2}(n\eta)}$ and
$G(\eta)=\frac{2c_{2}n\epsilon m^{2}{\rm
sinh}^{2}(n\eta)}{(\epsilon +{\rm sinh}^{2}(n\eta))^{2}}$ and
Case(ii):---
$\sigma(\eta)=\frac{A+B~\mbox{sech}^2(\eta)}{1+D~\mbox{sech}^2(\eta)}$
and
$G(\eta)=\frac{2c_2(AD-B)sech^2(\eta)tanh(\eta)}{(1+Dsech^2(\eta))^2}.$
These steep modulational profiles are specifically chosen because
they create tightly bound localized solitons, which are convenient
for the exact and numerical analysis alike.
\section{Chirped localized modes in $\mathcal{PT}$ symmetric optical media}
To start with, we have chosen the following form for complex field
    \be\label{3a2}\psi(z,\eta)=\rho(\eta)~e^{i(\phi(\eta)+\gamma z)},\ee
where $\rho$ and $\phi$ are the real functions of $\eta$ and
$\gamma$ is the propagation constant. The corresponding chirp is
given by $\delta\omega (\eta)=-\frac{\partial}{\partial
\eta}[(\phi (\eta)+\gamma z]=-\phi_{\eta} (\eta)$. Substituting
Eq. (\ref{3a2}) into Eq. (\ref{3a1}) and separating out the real
and imaginary parts of the equation, we arrive at the following
coupled equations in $\rho$ and $\phi$,
    \be\label{3a3}\rho_{\eta\eta}-2\gamma\rho-\phi_\eta^2\rho+2\sigma(\eta)\rho^3=0\ee
and
    \be\label{3a4}2\phi_\eta\rho_\eta+\phi_{\eta\eta}\rho-2G(\eta)\rho=0.\ee
To solve these coupled equations, we choose the following ansatz
    \be\label{3a03}\phi_\eta=c_1+c_2\rho^2.\ee
Hence, chirping is given as $\delta\omega(\eta)=-(c_1+c_2\rho^2)$,
where $c_1$ and $c_2$ denote the constant and nonlinear chirp
parameters, respectively. It means chirping of wave is directly
proportional to the intensity of wave. Using the ansatz Eq.
(\ref{3a03}), Eqs. (\ref{3a3}) and (\ref{3a4}) reduce to
    \bea\label{3a13}\rho_{\eta\eta}-(2\gamma+c_1^2)\rho-2c_1c_2\rho^3-2\sigma(\eta)\rho^3-c_2^2\rho^5=0,\\
    \label{3a14}G(\eta)=c_1\frac{\rho_\eta}{\rho}+2c_2~\rho\rho_\eta.\eea

The elliptic equation given by Eq. (\ref{3a13}) can be mapped to
$\phi^6$ field equation to obtain a variety of solutions such as
periodic, kink and solitary wave type solutions. In general, all
traveling wave solutions of Eq. (\ref{3a13}) can be expressed in a
generic form by means of the Weierstrass $\wp$ function. In this
work, we report only those soliton-like solutions for which gain
profile, given by Eq. (\ref{3a14}), remains localized. For
$c_2=0$, Eq. (\ref{3a13}) reduces to well known cubic elliptic
equation for which non-chirped soliton solutions can be easily
found. Here, we studied Eq. (\ref{3a13}) for different parameter
conditions and obtained chirped soliton solutions. The reported
solutions consist various soliton solutions like double-kink and
fractional-transform solitons that respect $\mathcal{PT}$
symmetry. For double-kink solitons, we consider the case $c_1=0$
because for non-zero values of $c_1$ the corresponding gain will
no longer be localized.

\subsection{Double-kink solitons}
Eq. (\ref{3a13}) admits double-kink solutions of the form
    \be\label{3a15}\rho(\eta)=\frac{m \sinh(n{\eta})}{\sqrt{\epsilon +\sinh^2(n{\eta})}},\ee
where $m=\left(\frac{-2\gamma}{2+c_2^2}\right)^{1/4}$,
$n=\sqrt{-2\gamma}$, $c_2=1$ and $c_1=0$. The choice $c_1=0$ has
been made in order to avoid the singularities in gain profile.
Here, $\epsilon$ is a free parameter which controls the width of
soliton solutions. The interesting double-kink feature of the
solution is prominent for large values of $\epsilon$. Now, for
$m,~n$ and $c_2$ to be real numbers, $\epsilon$ should be always
greater than $1$.
   For solution given in Eq. (\ref{3a15}), the gain will be of the
following form
    \be\label{3g1} G(\eta)=\frac{2c_2m^2n\epsilon sinh(n\eta)}{(\epsilon+sinh(n\eta))^2}
    \ee
and the chirping will be
    \be\label{3chirp1}\delta\omega(\eta)=-\frac{c_2~m^2\sinh^2(n\eta)}{\epsilon+\sinh^2(n\eta)}.\ee

   \begin{figure}[htbp]
            \includegraphics[width=3in]{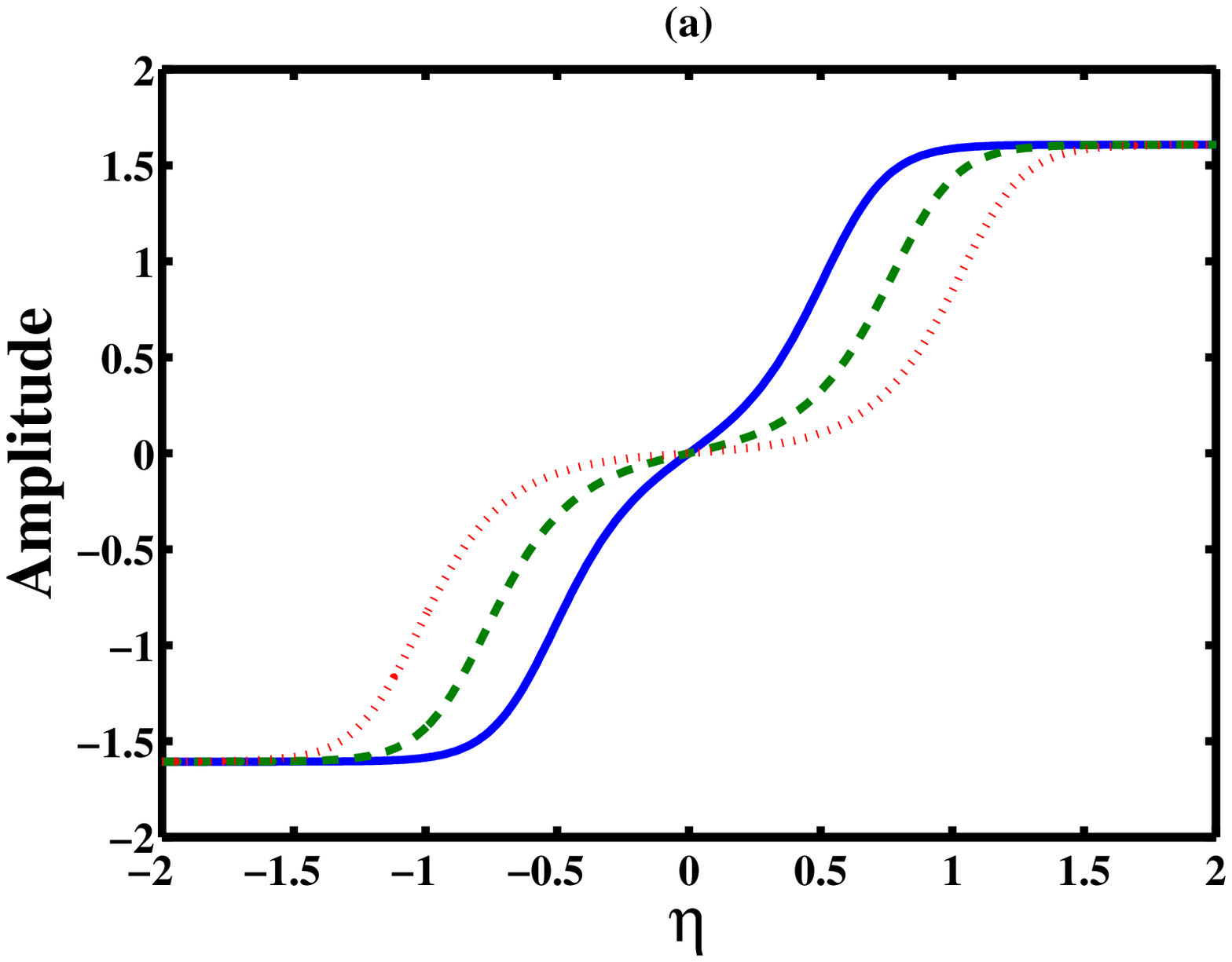}
            \includegraphics[width=3in]{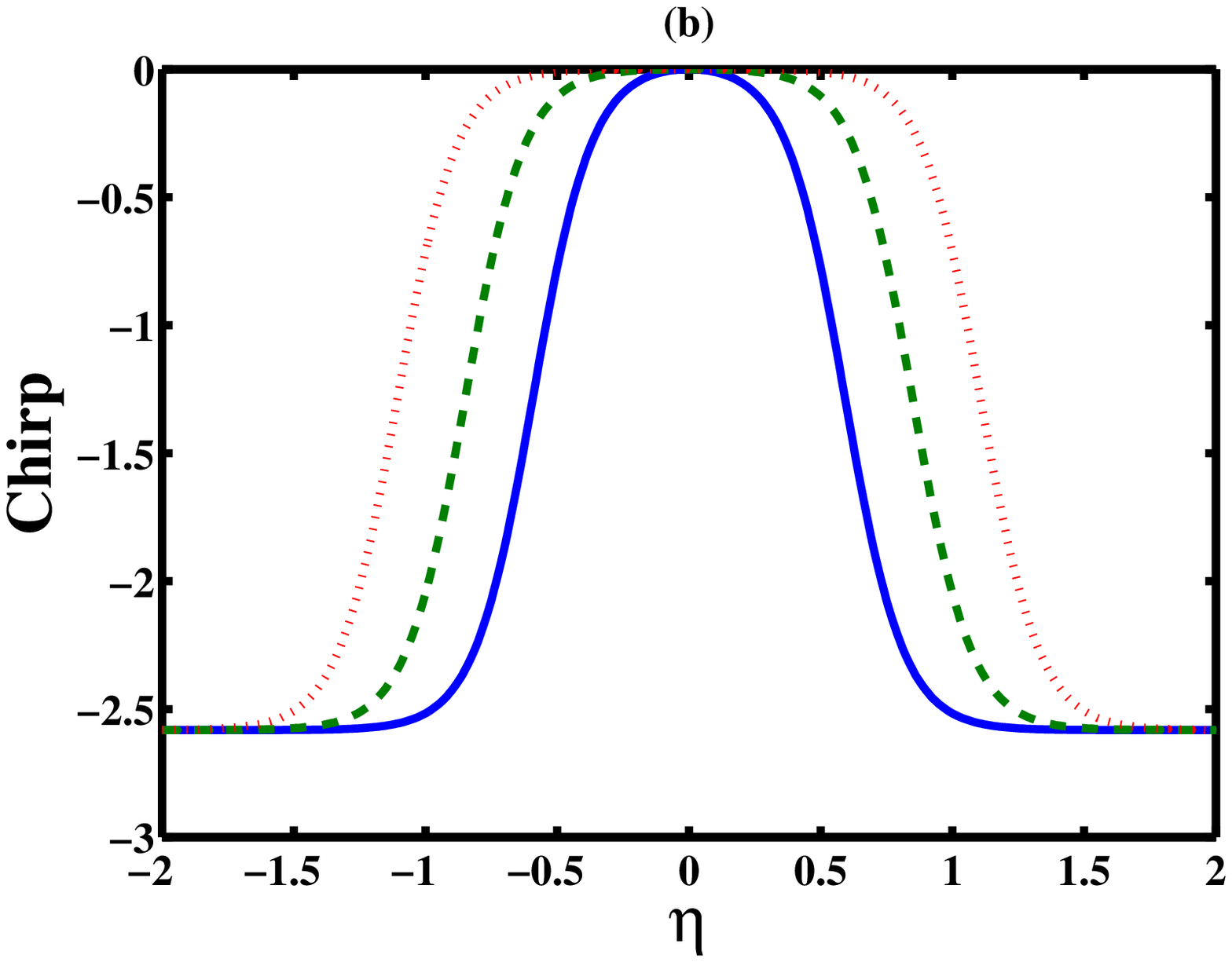}
            \centering
            \includegraphics[width=3in]{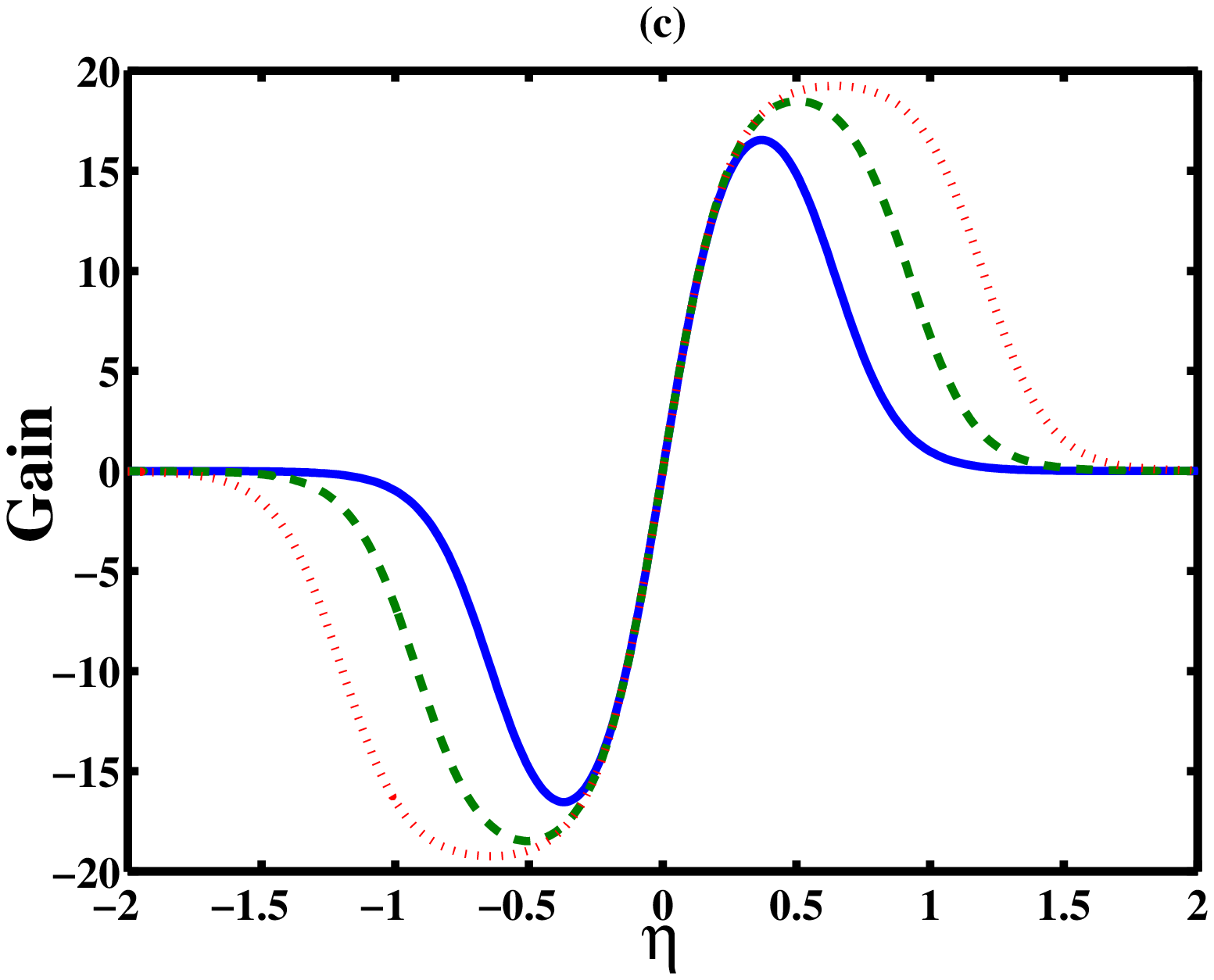}
        \caption{(a) Gain, (b) amplitude, and (c) chirp profiles for double-kink solitons for different values of
                    $\epsilon$, $\epsilon=50$ (thick line), $\epsilon=500$ (dashed line) and $\epsilon=5000$ (dotted line). The other parameter used in the plot is $ \gamma =-10$.}
        \label{fig_1.1}
    \end{figure}
\par The variation of gain, amplitude and chirp for different values of $\epsilon$ is
shown in Fig. \ref{fig_1.1}. The parameter used in the plot  is
$\gamma =-10$. One can point out that as the value of $\epsilon$
changes, there is a small change in the gain profile which in turn
have significant effect on the amplitude and chirp of double-kink
solution. Hence, one can observe that the double-kink feature of
the wave is more prominent for a gain medium with large values of
$\epsilon$, and different gain medium effects only the width of
the double-kink wave whereas the amplitude of the wave always
remains same. From the plot of chirp, it is clear that it has a
maximum at the center of the wave and saturates at some finite
value of $\eta$.

\section{Fractional-transform solitons} For the parametric
condition $c_1=0$, Eq. (\ref{3a13}) can be solved for very
interesting fractional-transform solitons
\cite{soloman1,soloman2}. To accomplish this, one can substitute
$\rho^2=y$ in Eq. (\ref{3a13}) to obtain the following equation:
    \be\label{3a015}y''+py^2+qy^3+ry+c_0=0,\ee
where $p=-b/2,~q=-4-2c_2^2,~r=-a-4\gamma$ and $c_0$ is integration
constant. For illustration purposes, here the values of $a,~b$ and
$c_0$ are chosen as $-84,~5.334$ and $-1$ respectively. In order
to find localized soliton solutions of Eq. (11), we use a
fractional transformation
    \be\label{3a16}y(\eta)=\frac{A+Bf^2(\eta)}{1+Df^2(\eta)},\ee
where the determinant $AD-B\neq 0$.

    \par
Our main aim is to study the localized solutions, we consider the
case where $f=\mbox{cn}(\eta,m)$ with modulus parameters $m=1$ and
$m=0$. By substituting Eq. (12) into Eq. (11) and equating the
coefficients of equal powers of Jacobian ${\rm cn}$ to zero, we
obtain the following consistency conditions for Jacobian elliptic
modulus $m=1$:
    \bea
    qA^3+pA^2+rA-c_0=0,\label{3a17}\\
    4(B-AD)+2pB+pA^2D+3A^2B\no\\+rB+2rAD-3Dc_0=0,\label{3a18}\\
    6(AD-B)+4D(AD-B)pB^2+2pBD\no\\+3AB^2+rAD^2+2rBD-3D^2c_0=0,\label{3a19}\\
    2D(B-AD)+pB^2D+qB^3+rBD^2-D^3c_0=0.\label{3a20}
    \eea
The set of Eqs. (\ref{3a17}) to (\ref{3a20}) can be solved
consistently for the unknown parameters $A,B,D$ and for a
particular value of $c_0$.
\subsection{Soliton solution}
The generic profile of the solution reads
    \be\label{3a21}y(\eta)=\frac{A+B~\mbox{sech}^2\eta}{1+D~\mbox{sech}^2\eta}.\ee
And, $\rho(t)$ can be written as
    \be\label{3a22}\rho(\eta)=\sqrt{\frac{A+B~\mbox{sech}^2\eta}{1+D~\mbox{sech}^2\eta}}.\ee

     \begin{figure}[htbp]
                \includegraphics[width=3in]{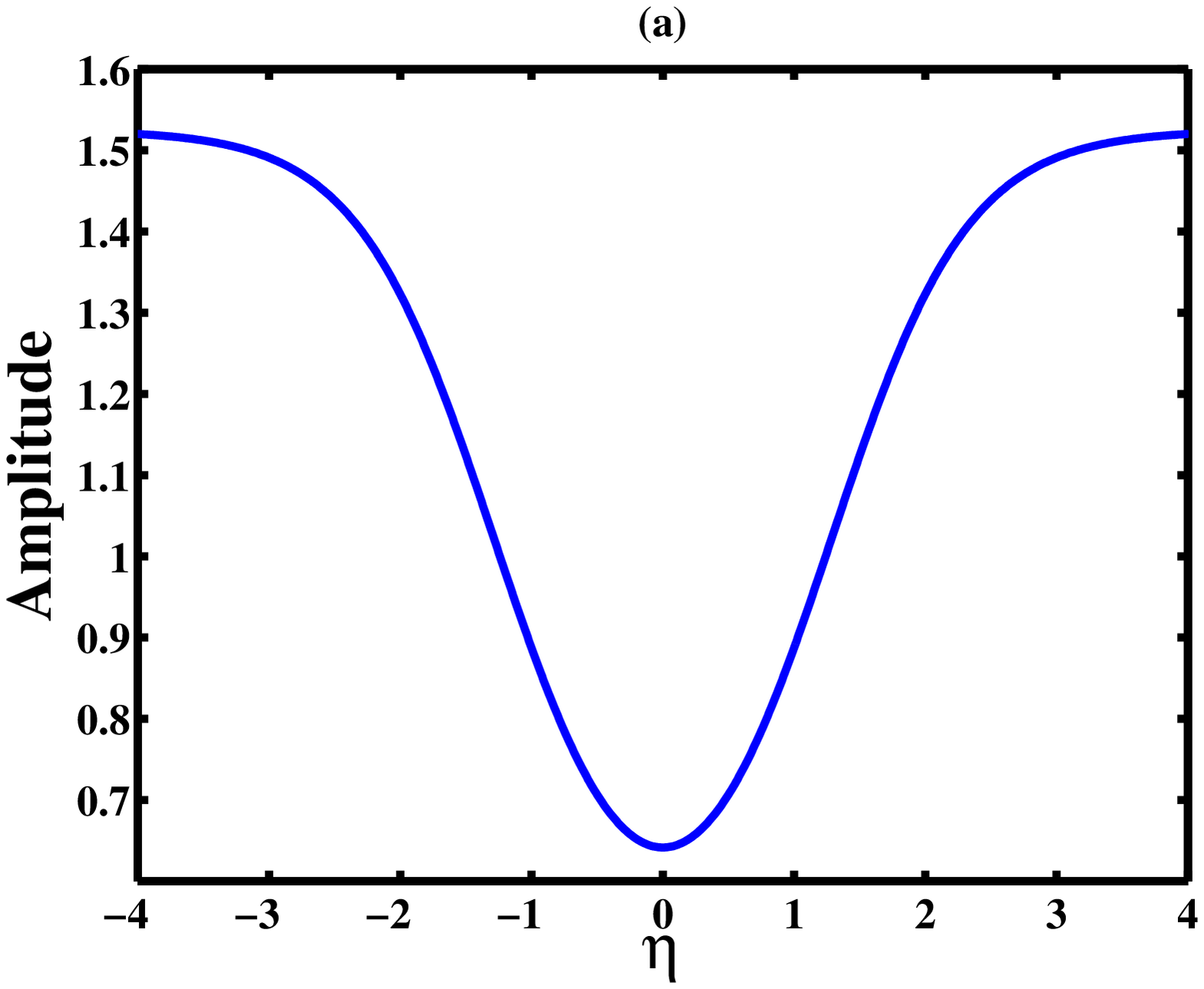}
                \includegraphics[width=3in]{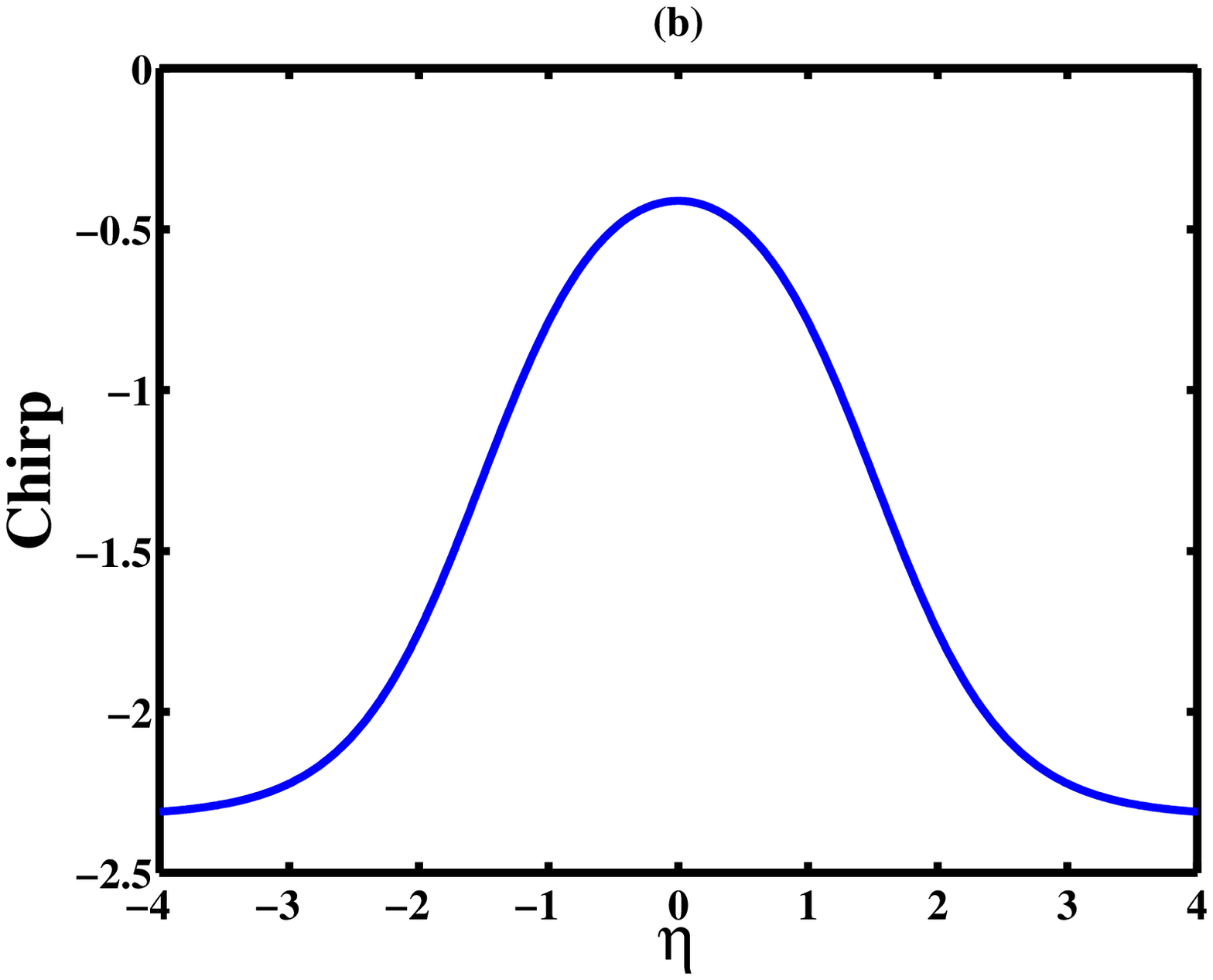}
                 \centering
                 \includegraphics[width=3in]{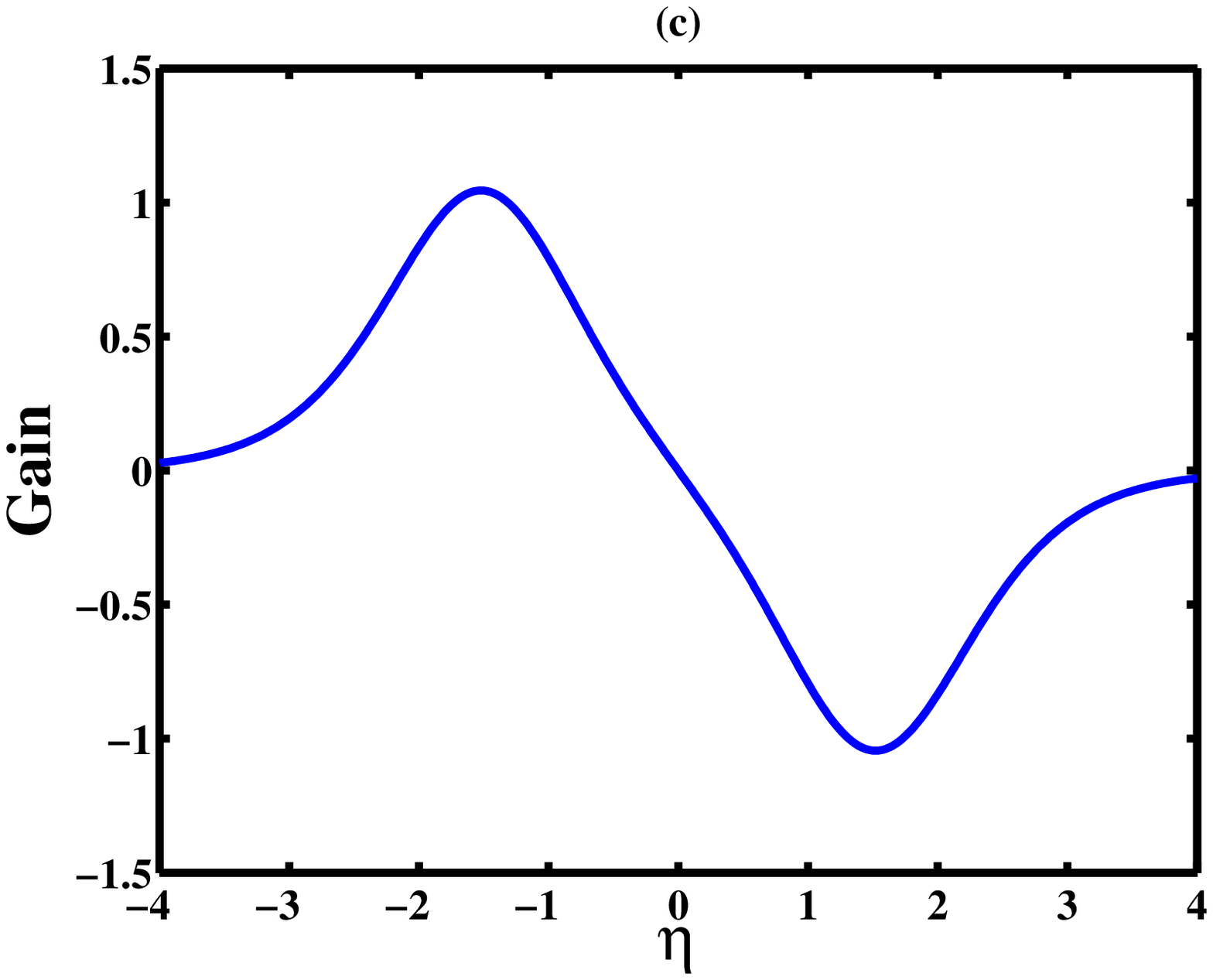}
            \caption{(a) Amplitude, (b) Chirp and (c) Gain profiles for fractional-transform dark solitons for $c_2=1$, $c_0=-1$ and $\gamma =10$..}
            \label{fig_1.2}
        \end{figure}
Since the analytical form of solution is known, a simple
maxima-minima analysis can be done to distinguish parameter
regimes supporting dark and bright soliton solutions. In this
case, when $AD<B$ one gets a bright soliton, whereas if $AD>B$
then a dark soliton exists. \par For soliton solution given in Eq.
(\ref{3a22}), the chirping is given by
    \be\label{3a022}\delta\omega(\eta)=-c_2 \frac{A+B~\mbox{sech}^2\eta}{1+D~\mbox{sech}^2\eta}, \ee
The amplitude, chirp and gain profiles for fractional-transform
soliton are shown in Figs. \ref{fig_1.2}(a), \ref{fig_1.2}(b) and
\ref{fig_1.2}(c)  respectively for $ c_0=-1,~c_2 = 1~
\mbox{and}~\gamma=10 $. For these values, the various unknown
parameters have been found to be $A=2.56124,~B=0$ and $D=5.12248$.
Here, solution is of the form of dark soliton and has a small
amplitude over a finite background. For this case, chirping is
maximum at the center of the wave and is dominant away from the
center.

The corresponding gain will be of the following form
    \bea\label{3a23}G(\eta)=\frac{2c_2(AD-B)sech^2(\eta)tanh(\eta)}{(1+Dsech^2(\eta))^2}.\eea

\subsection{Trigonometric solution}
For the Jacobian elliptic modulus $m=0$, we obtain a very
interesting trigonometric solution. Following are the consistency
conditions we obtain for this Jacobian elliptic modulus:

\bea
    qA^3+pA^2+rA-(c_0+1)=0,\label{3a24}\\
    4(AD-B)(2+3D)+2pB+pA^2D+3qA^2B\no\\+rB+2rAD-3Dc_0=0,\label{3a25}\\
    4D(AD-B)+pB^2+2pBD\no\\+3AB^2q+rAD^2+2rBD-3D^2c_0=0,\label{3a26}\\
    pB^2D+qB^3+rBD^2-D^3c_0=0.\label{3a27}
    \eea

    \be\label{3a29}y(\eta)=\frac{A+B~\mbox{cos}^2\eta}{1+D~\mbox{cos}^2\eta}.\ee
    And, $\rho(t)$ can be written as
        \be\label{3a30}\rho(\eta)=\sqrt{\frac{A+B~\mbox{cos}^2\eta}{1+D~\mbox{cos}^2\eta}}.\ee

Chirp is given by
 \be\label{3a031}\delta\omega(\eta)=-c_2 \frac{A+B~\mbox{cos}^2\eta}{1+D~\mbox{cos}^2\eta}, \ee

The corresponding gain will be of the following form
\bea\label{3a32}G(\eta)=\frac{2c_2(AD-B)cos(\eta)sin(\eta)}{(1+Dcos^2(\eta))^2}.\eea

\begin{figure}[htbp]
                \includegraphics[width=3in]{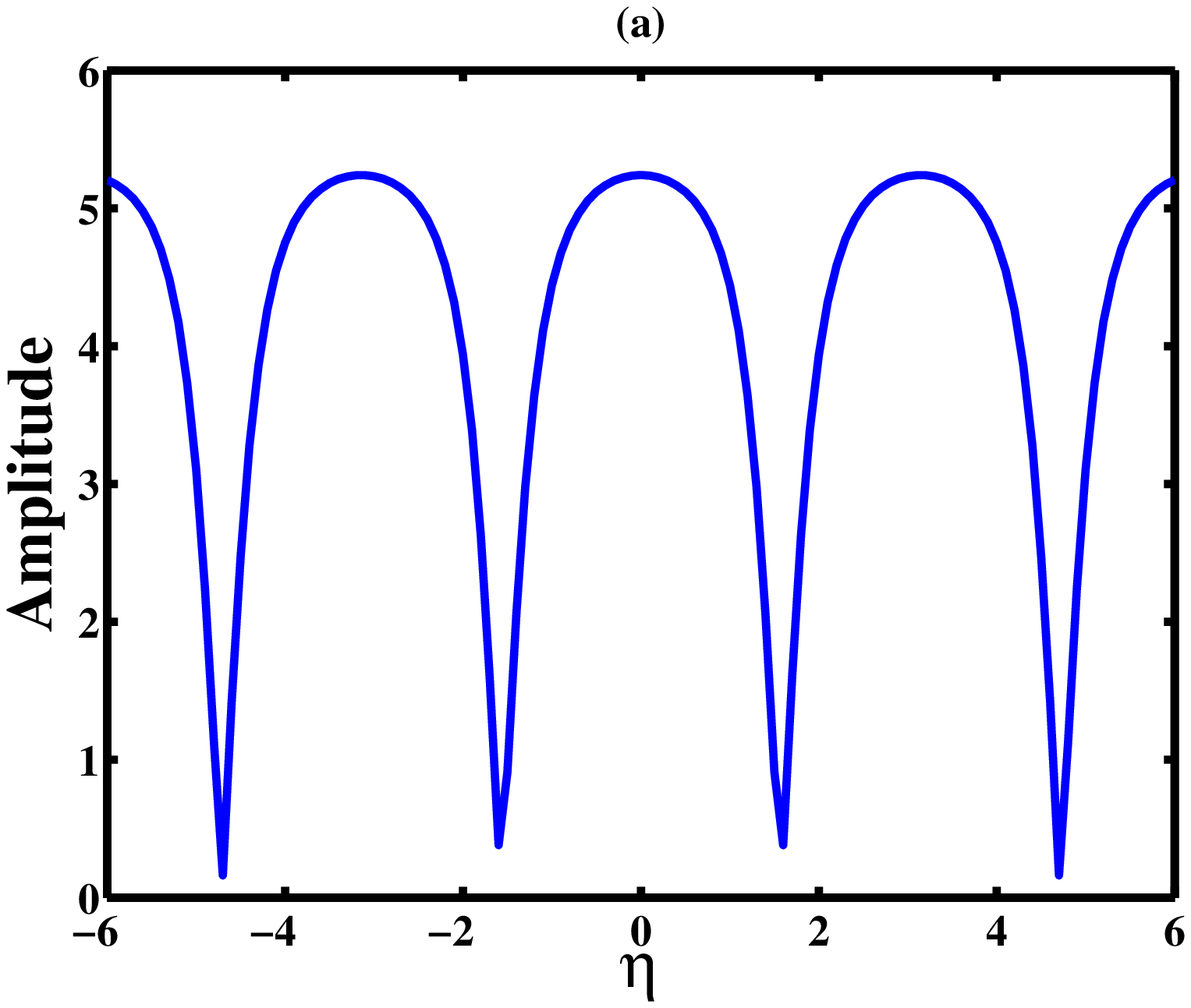}
                \includegraphics[width=3in]{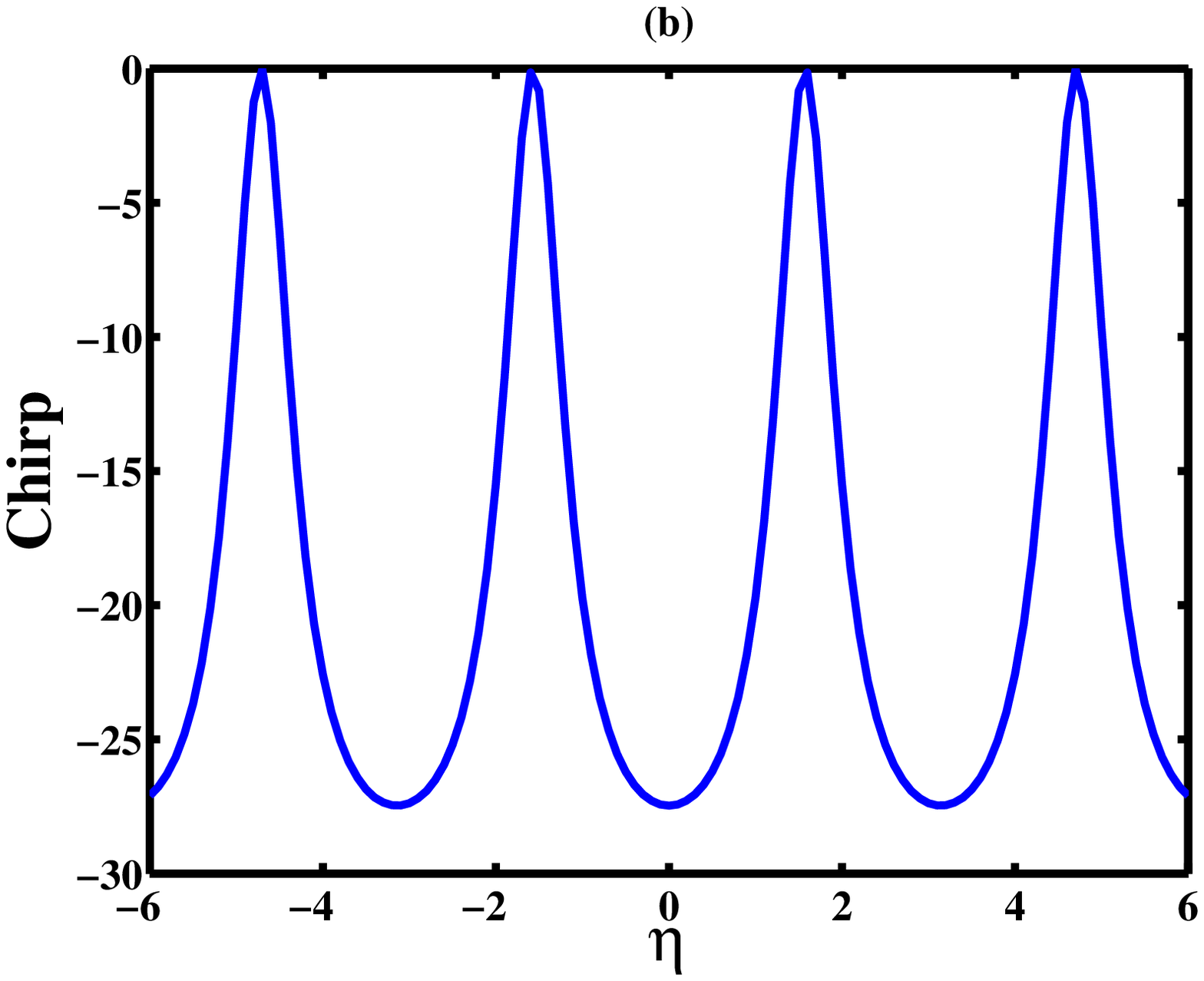}
                \centering
                \includegraphics[width=3in]{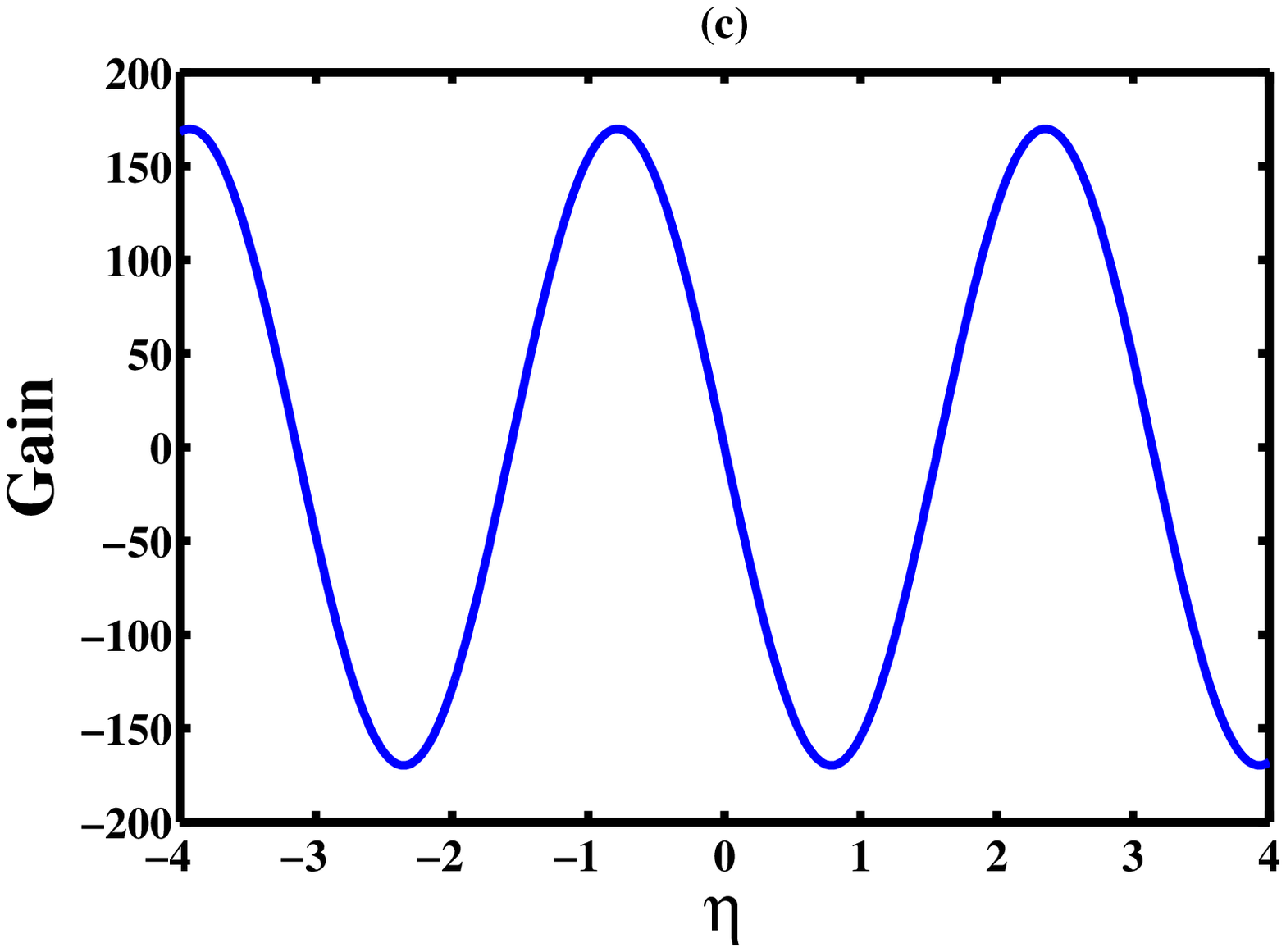}
            \caption{(a) Amplitude, (b) Chirp and (c) Gain profiles for fractional-transform dark solitons for $c_2=1$, $c_0=-1$ and $\gamma =10$.}
            \label{fig_1.3}
        \end{figure}

\section{Stability of the localized modes in $\mathcal{PT}$ symmetric optical media} In this section,
we study the stability of the $\mathcal{PT}$ symmetric periodic
and hyperbolic fractional-transform solitons under the evolution
of Eq. (1), numerically using the Crank-Nicolson finite-difference
method, which is unconditionally stable. Figures 4(a) and 4(b),
depict the perturbed solution with the initial condition
$\psi(\eta,z=0)=\psi(\eta,z=0)+\e$, where $\e$ is a function which
assumes a random value at each point. It is found that the maximum
value of $\e$ is 10\% of the peak value of the intensity profile.
One can clearly see from Fig. 4(a) and Fig. 4(b), which are the
z-evolution of  periodic and hyperbolic fractional-transform
solitons, that perturbation doesn't destabilize the solutions. We
have checked addition of random noise up to 30\% of the peak value
of the intensity profile; the solitons are found to be quite
stable. In the case of periodic soliton, as evidenced from Fig.4
(a), the fractional-transform soliton is found to be quite stable,
although the peak of the intensity oscillates. The step size
$d\eta$ and $dz$ were taken as $0.01$ and $0.0001$, repectively,
in these simulations.

\begin{figure}
 \includegraphics[width=3in]{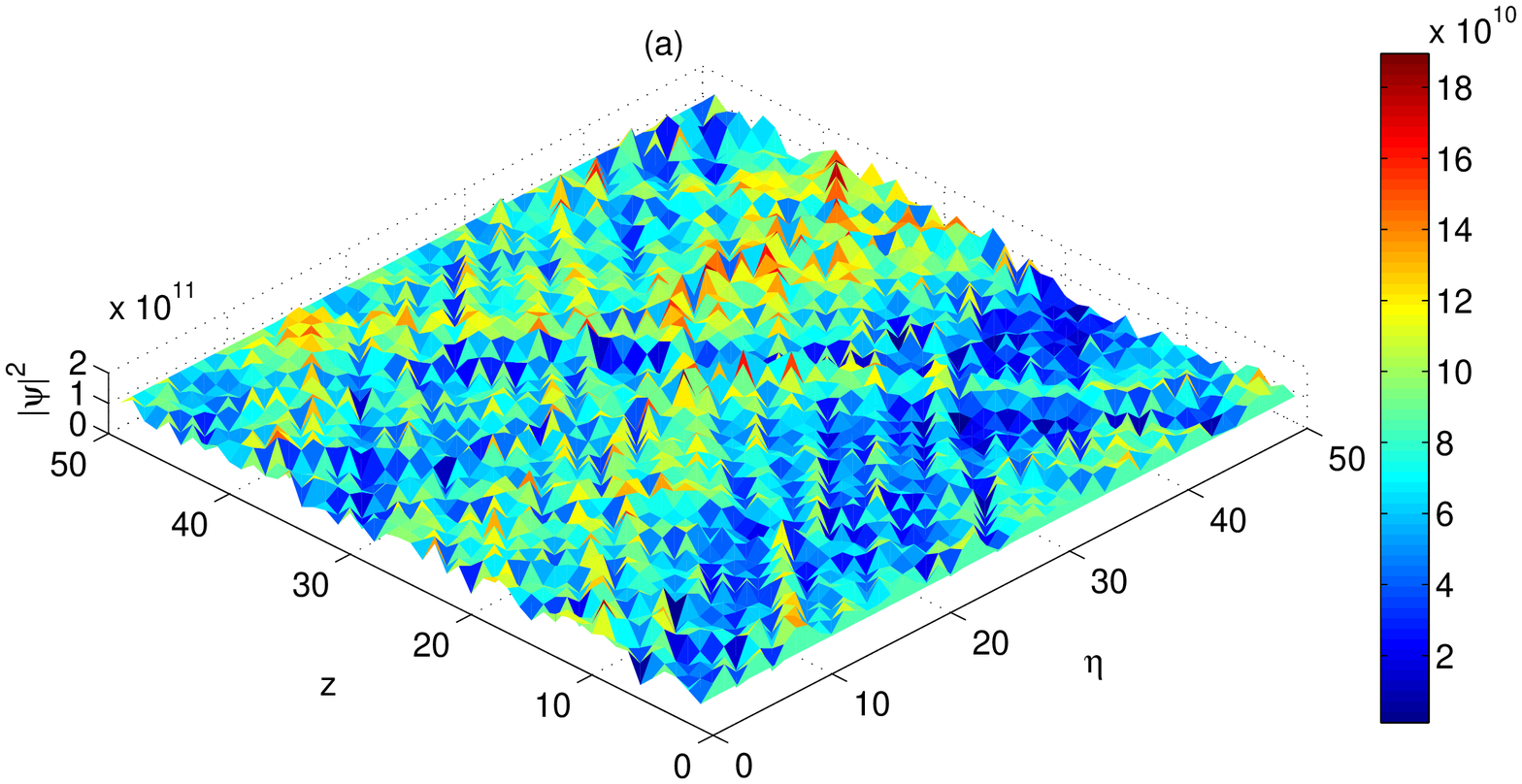}\\
\includegraphics[width=3in]{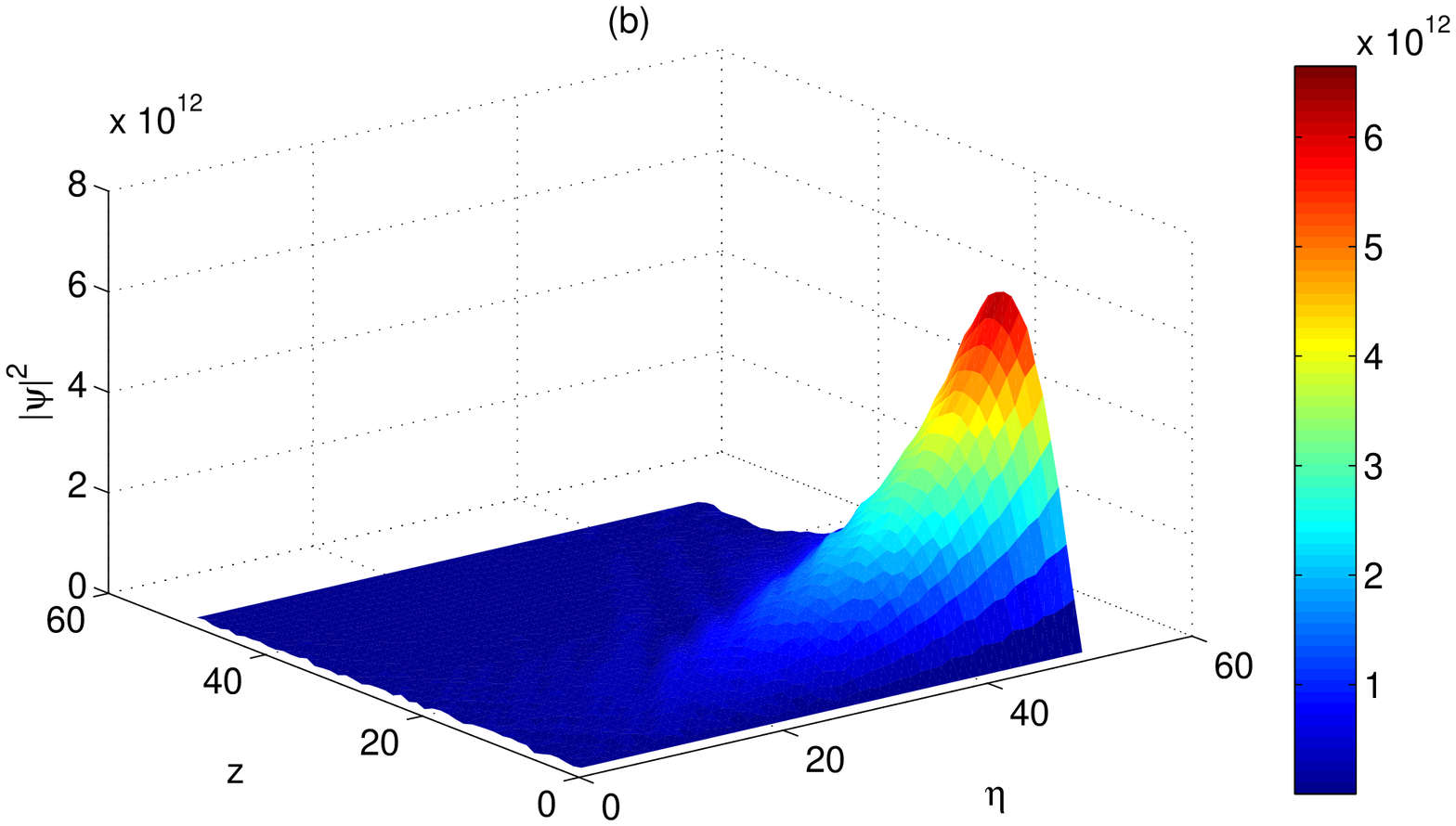}
\caption{Intensity evolution of nonlinear mode for  (a)
fractional-transform trigonometric solution and (b)
fractional-transform soliton solution.}
\end{figure}

\section{Conclusion}
In conclusion, we have demonstrated the existence of exact
unbreakable $\mathcal{PT}$ symmetric chirped double-kink solitons
and fractional-transform solitons in nonlinear optical media for
the propagation of a laser beam. We have exemplified this
phenomenon of unbreakable $\mathcal{PT}$ symmetry for two specific
cases of nonlinearity function and gain-loss profiles. Toward the
end, we have conducted numerical stability experiments of the
$\mathcal{PT}$ symmetric localized excitations---periodic and
hyperbolic fractional-transform solitons---using a semi-implicit
Crank-Nicolson finite difference algorithm and found that they are
quite stable, against finite perturbations. Even though there can
be other exact solutions found out for this dynamical system, all
of them may not satisfy the $\mathcal{PT}$ symmetry.

\section*{ACKNOWLEDGEMENT}
T.S.R. acknowledges support from the DST, Government of India,
through Fast Track Project Ref. No.--SR/FTP/PS-132/2012, during
the course of this work.

\end{document}